\documentclass[11pt,letterpaper]{article}
\pdfoutput = 1
\usepackage     {jheppub}
\usepackage     [mathscr]{eucal}
\usepackage     {bm}
\bibpunct	{[}{]}{,}{n}{}{;}

\def\T		{\widehat T}
\def\P		{P}

\def\del {\nabla}

\def\clr#1{{{#1}}}

\def\tocsqueeze	{\vspace*{-3.7pt}}

\title          { How big are the smallest drops of quark-gluon plasma?}
\author[a]      {Paul~M.~Chesler,}
\affiliation[a]	{Department of Physics, Harvard University, Cambridge MA 02138, USA}
\emailAdd	{pchesler@physics.harvard.edu}

\abstract
    {%
    Using holographic duality, we present results for both head-on and off-center collisions of 
    Gaussian shock waves in strongly coupled 
    $\mathcal N = 4$ supersymmetric Yang-Mills theory.  The shock waves 
    superficially resemble Lorentz contracted colliding protons.  The collisions results in the formation of a plasma 
    whose evolution is well described by viscous hydrodynamics.  The size of the produced droplet is 
    $R \sim 1/T_{\rm eff}$ where $T_{\rm eff}$ is the effective temperature, which is the characteristic 
    microscopic scale in strongly coupled plasma.
    These results demonstrate the applicability of hydrodynamics to microscopically small systems and bolster the notion
    that hydrodynamics can be applied to heavy-light ion collisions as well as some proton-proton collisions.
    }

\keywords	{general relativity, gauge-gravity duality, 
		 quark-gluon plasma}
\begin{document}

\advance\textheight 55pt
\maketitle
\thispagestyle{empty}
\advance\textheight -55pt

\addtocontents	{toc}{\tocsqueeze}
\addtocontents	{toc}{\tocsqueeze}

\section{Introduction}

Hydrodynamics is a long wavelength late time effective description of the transport
of conserved degrees of freedom.  Hydrodynamics lacks many of the excitations of
the underlying quantum theory (\textit{i.e.} those excitations which are not conserved).  
Other than their effect on transport coefficients, 
it is consistent to neglect such excitations in the macroscopic limit since they attenuate over 
microscopic scales.

A striking application of hydrodynamics is the modeling of heavy-light ion collisions and proton-proton collisions at RHIC and the LHC, where signatures of collective flow have been observed 
\cite{Chatrchyan:2013nka,Abelev:2012ola,Aad:2012gla,Adare:2013piz,Adare:2015ctn, Aad:2015gqa}
which are consistent with hydrodynamic evolution of quark-gluon plasma 
\cite{Bozek:2011if,Bozek:2012gr,Bzdak:2013zma,Schenke:2014zha,Habich:2015rtj}.  
The success of hydrodynamics 
is surprising since at experimentally accessible energies, microscopic scales such at the mean free
path are probably not too different from the system size.  This raises several interesting questions.
For such tiny systems is it theoretically consistent to neglect nonhydrodynamic degrees of freedom?  
Does the presence of large gradients inherent to small systems excite nonhydrodynamic modes or otherwise spoil
the hydrodynamic gradient expansion?  How big are the smallest drops of matter which can meaningfully 
be identified as liquids?  Are we deluding ourselves into thinking hydrodynamics can apply to systems as small as a proton?

Since microscopic length and time scales typically becomes smaller
in the limit of strong coupling, it is natural to expect the domain of utility of hydrodynamics 
to be largest at strong coupling.  However, strongly coupled dynamics
in QCD are notoriously difficult to study.  It is therefore of great utility to have a toy model of 
strongly coupled dynamics --- which accounts for both hydrodynamic and nonhydrodynamic excitations 
--- where one can answer the above questions in a controlled and systematic
setting.  

Holographic duality \cite{Maldacena:1997re} 
maps the dynamics of certain strongly coupled non-Abelian 
gauge theories onto the dynamics of classical gravity in higher dimensions.  The process of 
quark-gluon plasma formation maps onto the process of gravitational collapse and 
black hole formation, with the ring down of the black hole encoding the relaxation of the plasma
to a hydrodynamic description.  The dual gravity calculation is microscopically complete, 
meaning it contains both hydrodynamic and nonhydrodynamic excitations.  In other words,
all stages of evolution --- from far-from-equilibrium dynamics to hydrodynamics --- are all encoded 
Einstein's equations.  Since Einstein's equations can be solved numerically, holographic duality 
provides a unique arena where one can systematically test the domain of utility of hydrodynamics. 
We shall therefore use holographic duality to test hydrodynamics in extreme conditions.

The simplest theory with a dual gravity description is $\mathcal N = 4$ supersymmetric Yang-Mills 
theory (SYM), which is dual to gravity in ten dimensional AdS$_5$$\times$$S^5$ spacetime.
A simple model of quark-gluon plasma production in SYM is the collision of gravitational 
shock waves 
\cite{
  Grumiller:2008va,
  Albacete:2008vs,
  Chesler:2010bi,
  Casalderrey-Solana:2013aba,
  Chesler:2015fpa,
  Chesler:2015bba,
  Chesler:2015wra}, 
which can result in the formation of a black hole in AdS$_5$$\times$$S^5$.  The shock's waveform,
which is not fixed by Einstein's equations, determines the expectation value of the stress tensor 
in the dual field theory.  Previously, in \cite{Chesler:2015bba} we considered the collision of a localized shock wave 
with a planar shock wave, which in the dual field theory resembled a
proton-nucleus collision.  There it was found that the collision resulted in a microscopically small droplet 
plasma being produced, whose evolution was well described by viscous hydrodynamics.
Here we shall extend our 
work to include the collisions 
which resemble proton-proton  collisions.
In particular, we shall consider the collision two shock waves moving in the $\pm z$ directions at the speed of light
with expectation value of the SYM energy density
\begin{equation}
\label{eq:singleshock}
\langle T^{00}_{\rm single \ shock} \rangle = \frac{N_{\rm c}^2}{2 \pi^2} \mu^3 
\exp \left [-{\textstyle \frac{1}{2}}( \bm x_\perp \pm \bm b/2)^2/\sigma^2 \right]  \delta_w(z\mp t),
\end{equation}
where $N_{\rm c}$ is the number of colors, $\mu$ is an  energy scale, $\delta_w$ is a smeared $\delta$ function, and
$\bm x_\perp =\{x,y\}$ are the coordinates transverse to the collision axis.
Hence our shock waves are Gaussians with transverse width $\sigma$, impact parameter 
$\bm b$, and total energy
\begin{equation}
\label{eq:shockenergy}
E_{\rm single \ shock} = \int d^3 x \, \langle  T^{00}_{\rm single \ shock} \rangle  = 
\frac{N_c^2}{2 \pi^2} (2 \pi \sigma^2) \mu^3.
\end{equation}
Superficially at least, the shocks resemble Lorentz contracted colliding protons.  We shall refer to the shocks 
as ``protons" with the quotes to emphasize to the reader than we are not colliding bound states
in QCD but rather caricatures in SYM.  The precollision bulk geometry contains a trapped surface and
the collision results in the formation of a black hole. We
numerically solve the Einstein equations for the geometry after the collision and report on the evolution
of the expectation value of the SYM stress tensor $\langle T^{\mu \nu} \rangle$  and test the validity of
hydrodynamics.  

In strongly coupled SYM the hydrodynamic gradient expansion 
is an expansion in powers of $1/(\ell T_{\rm eff})$ with $\ell$ the characteristic scale over
which the stress tensor varies and $T_{\rm eff}$ the effective temperature.  Therefore, in
strongly coupled SYM the scale $1/T_{\rm eff}$ plays the role of a mean free path.  We focus on the low 
energy limit where $T_{\rm eff}$ is small and $1/T_{\rm eff}$ is large.
Without loss of generality, in the results presented below we measure all quantities in units of $\mu = 1$.
We choose shock parameters
\begin{align}
\label{eq:shockparms}
& \sigma = 3,&
&\bm b = 0,& 
&\bm b = 3 \, \hat x.&
\end{align}
We shall refer to the $\bm b = 0$ and $\bm b = 3 \hat x$ collisions simply as ``head-on"
and ``off-center," respectively.
Both 
collisions result in the formation of a microscopically small droplet plasma of size $R \sim 1/T_{\rm eff}$,
whose evolution is well described by viscous hydrodynamics.
These results demonstrate that hydrodynamics can work in microscopically small systems and 
bolster the notion that the collisional debris in heavy-light ion collisions and proton-proton collisions
can be modeled using hydrodynamics.  

An outline of the rest of the paper is as follows.  In Sec.~\ref{sec:hydrodynamization} we construct a 
simple test to see whether the evolution of $\langle T^{\mu \nu} \rangle$ is governed by hydrodynamics.
In Sec.~\ref{sec:gravdesc} we outline the gravitational formulation of the problem.  In Sec.~\ref{sec:results}
we present our results for the evolution of $\langle T^{\mu \nu} \rangle$ and in Sec.~\ref{sec:discuss} we discuss
our results and generalizations to confining theories.  Readers not interesting the details of the gravitational
calculation can skip Sec.~\ref{sec:gravdesc}.

\section{A litmus test for hydrodynamic evolution of $\langle T^{\mu \nu} \rangle$}
\label{sec:hydrodynamization}

With the gravitational calculation presented below we shall obtain the exact expectation value of the 
SYM stress tensor $\langle T^{\mu \nu} \rangle$.  At time and length scales $\gg$ than microscopic scales 
the evolution of $\langle T^{\mu \nu} \rangle$ must be governed by 
hydrodynamics.  
This means that 
\begin{equation}
\label{eq:hydrodyanmization}
\langle T^{\mu \nu}\rangle \approx T^{\mu \nu}_{\rm hydro}(\epsilon, u^\mu),
\end{equation}
where $T^{\mu \nu}_{\rm hydro}(\epsilon,u^\mu)$ is given in terms of the proper energy density $\epsilon$
and the fluid velocity $u^\mu$ via the hydrodynamic constitutive relations.  When (\ref{eq:hydrodyanmization}) 
is satisfied we shall say that the system has \textit{hydrodynamized}.  In this section our goal is develop a 
simple test, which requires only $\langle T^{\mu \nu} \rangle$, to see whether the system has hydrodynamized.  
We shall not give a through review of hydrodynamics here and instead shall only state the salient 
features of hydrodynamics needed for our analysis.  For more detailed discussions of relativistic hydrodynamics
we refer the reader to \cite{Baier:2007ix,Romatschke:2009im}.

We use the mostly positive Minkowski space metric $\eta^{\mu \nu} = {\rm diag}(-1,1,1,1)$ and 
define the fluid velocity to be the timelike ($u^0 > 0$) normalized ($u_\mu u^\mu = -1$) 
eigenvector of $T^{\mu \nu}_{\rm hydro}$ with $-\epsilon$ the associated eigenvalue, 
\begin{equation}
\label{eq:veldef0}
T^{\mu \nu}_{\rm hydro} u_{\nu} = - \epsilon u^\mu.
\end{equation}
With these conventions, at second order in gradients 
the hydrodynamic constitutive relations in strongly coupled SYM read \cite{Bhattacharyya:2008jc,Baier:2007ix}
\begin{equation}
\label{eq:hydrocons}
T^{\mu \nu}_{\rm hydro} = p \, \eta^{\mu \nu} + (\epsilon + p) u^\mu u^\nu + \mathcal T^{\mu \nu}_{(1)} + \mathcal T^{\mu \nu}_{(2)},
\end{equation}
where  $p$ is the pressure
given by the conformal equation of state
\begin{equation}
\label{eq:eos}
p = \frac{\epsilon}{3},
\end{equation}
and $\mathcal T^{\mu \nu}_{(1)}$ and
$\mathcal T^{\mu \nu}_{(2)}$ are the first and second order gradient corrections, respectively.
Explicitly, 
\begin{align}
&\mathcal T^{\mu \nu}_{(1)} = - \eta \sigma^{\mu \nu},&
&\mathcal T^{\mu \nu}_{(2)} = \eta \tau_{\Pi} \left [ ^{\langle} {D\sigma^{\mu \nu}}^{\rangle} + {\textstyle \frac{1}{3}} \sigma^{\mu \nu} (\partial \cdot u)\right]
+ \lambda_1 \sigma^{\langle \mu}_{\ \lambda}\sigma^{\nu \rangle \lambda}
+ \lambda_2 \sigma^{\langle \mu}_{\ \lambda}\Omega^{\nu \rangle \lambda},
\end{align}
where $D \equiv u \cdot \partial$, and the shear $\sigma^{\mu \nu}$ and vorticity $\Omega^{\mu \nu}$ tensors are 
\begin{align}
&\sigma^{\mu \nu} \equiv 2 ^{\langle} \partial^{\mu} u^{\nu \rangle}, &
&\Omega^{\mu \nu} \equiv \frac{1}{2} \P^{\mu \alpha} \P^{\nu \beta} (\partial_{\alpha} u_\beta -\partial_{\beta} u_\alpha),&
\end{align}
with the transverse projector
\begin{equation}
P^{\mu \nu} \equiv g^{\mu \nu} + u^\mu u^\nu.
\end{equation}
Note $u_\mu P^{\mu \nu} = 0$ since $u_\mu u^\mu = - 1$.
For any tensor $A^{\mu \nu}$ the bracketed tensor $A^{\langle \mu \nu \rangle}$ is defined 
to be the symmetric ($A^{\langle \mu \nu \rangle} = A^{\langle \nu \mu \rangle}$), traceless ($\eta_{\mu \nu} A^{\langle \mu \nu \rangle} = 0$) and 
transverse ($u_\mu A^{\langle \mu \nu \rangle} = 0$) component of $A^{\mu \nu}$, meaning  
\begin{equation}
A^{\langle \mu \nu \rangle} \equiv \frac{1}{2} \P^{\mu \alpha} \P^{\nu \beta} (A_{\alpha \beta} + A_{\beta \alpha}) - \frac{1}{3} \P^{\mu \nu} \P^{\alpha \beta} A_{\alpha \beta}.
\end{equation}
In strongly coupled SYM the shear viscosity $\eta$, second order transport coefficients 
$\tau_{\Pi}$, $\lambda_1$, $\lambda_2$ and effective temperature $T_{\rm eff}$
read \cite{Policastro:2001yc,Bhattacharyya:2008jc,Baier:2007ix}
\begin{align}
\label{eq:transcoeffs}
&\eta = \frac{1}{3 \pi T_{\rm eff}} \epsilon,&
&\tau_{\Pi} = \frac{2 - \log 2}{2 \pi T_{\rm eff}},&
&\lambda_1 = \frac{\eta}{2 \pi T_{\rm eff}},
&\lambda_2 = -\frac{\eta \log 2}{\pi T_{\rm eff}},&
&T_{\rm eff} = \left ( \frac{8 \epsilon}{3 \pi^2 N_{\rm c}^2} \right )^{1/4}.&
\end{align}

The hydrodynamic equations of motion for $\epsilon$ and $u^{\mu}$ 
are given by the energy-momentum conservation equation
$\partial_\mu T^{\mu \nu}_{\rm hydro} = 0$.  One option to test for hydrodynamization
is to construct initial data for $\epsilon$ and $u^\mu$ 
and then to evolve $\epsilon$ and $u^\mu$  forward in time via the hydrodynamic 
equations of motion.  One can then reconstruct $T^{\mu \nu}_{\rm hydro}$ and  
compare it to the full stress $\langle T^{\mu \nu} \rangle$ to test the applicability of hydrodynamics.

%

However, given that we will have the full expectation value $\langle T^{\mu \nu} \rangle$,
solving the hydrodynamic equations of motion is unnecessary.
Consider the eigenvalues 
$p_{(\lambda)}$ and associated eigenvectors $e_{(\lambda)}^\mu$ of $\langle T^{\mu}_{\ \nu} \rangle$
\begin{equation}
\langle T^{\mu}_{\ \nu} \rangle \, e^\nu_{(\lambda)} = p_{(\lambda)} \, e^\mu_{(\lambda)} \,,
\label{eq:veldef}
\end{equation}
with no sum over $\lambda$ implied. If the system has hydrodynamized, then 
$\langle T^{\mu}_{\ \nu} \rangle$ should have one timelike eigenvector $e_{(0)}^\mu$, which according to 
(\ref{eq:hydrodyanmization}) and (\ref{eq:veldef0}), is just  the fluid velocity.
We therefore \textit{define} the fluid velocity and proper energy to be
\begin{align}
\label{eq:fluidveldef}
&u^\mu \equiv e_{(0)}^\mu,&
&\epsilon \equiv - p_{(0)}.
\end{align}
With the complete spacetime dependence of the flow field $u^\mu$ and energy density $\epsilon$ determined from the exact stress, we can construct  $T^{\mu \nu}_{\rm hydro}$
using Eqs.~(\ref{eq:hydrocons})--(\ref{eq:transcoeffs}).  

It will be useful to quantify the degree in which the system has hydrodynamized.
To this end we define the dimensionless residual measure
\begin{equation}
\label{eq:Delta}
\Delta(t,\bm x) \equiv \underset{t' \ge t}{\max} \left [ \frac{ \big | \big | \langle T^{\mu \nu}(t',\bm x) \rangle - T^{\mu \nu}_{\rm hydro}(t',\bm x) \big | \big |}{p(t',\bm x)} \right ],
\end{equation}
where $||A^{\mu \nu}||$ denotes the $L_2$ norm of $A^{\mu \nu}$, which for symmetric $A^{\mu \nu}$
is the magnitude of the largest magnitude eigenvalue of 
$A^{\mu \nu}$.
Why do we take the max in (\ref{eq:Delta})?
Note that in general $\big | \big | \langle T^{\mu \nu} \rangle- T^{\mu \nu}_{\rm hydro} \big | \big |$ is nonmonotonic
in time due to nonhydrodynamic modes, such as quasinormal modes, oscillating in time while decaying.   Taking the max in (\ref{eq:Delta}) ameliorates the effect of the oscillations and aids in avoiding the
false identification of hydrodynamic evolution if $\big | \big | \langle T^{\mu \nu} \rangle- T^{\mu \nu}_{\rm hydro} \big | \big |$ is momentarily small.
If $\Delta(t,\bm x) \ll 1$, then $\langle T^{\mu \nu}\rangle$ is well described by the hydrodynamic constitutive relations at $\bm x$ at time $t$ and all 
future times.  

We therefore have achieved our goal of quantifying hydrodynamization purely in 
terms of $\langle T^{\mu \nu} \rangle$. 
To reiterate the strategy, one first extracts the proper energy density $\epsilon$
and fluid velocity $u^\mu$ from $\langle T^{\mu \nu} \rangle$.  One then uses the 
hydrodynamic constitutive relations to compute $T^{\mu \nu}_{\rm hydro}$.
Finally, one then constructs the hydrodynamic residual $\Delta$ and looks for regions of 
spacetime where $\Delta \ll 1$ to identify hydrodynamic behavior.

\section{Gravitational description}
\label{sec:gravdesc}

According to holographic duality, the evolution of the expectation value of the stress
tensor in strongly coupled SYM is encoded in the gravitational 
field in asymptotically AdS$_5$$\times$$S^5$ spacetime.   Einstein's equations are 
consistent with no dynamics on the $S^5$.  We shall make this assumption for our numerical 
analysis and revisit it in the Discussion section below.  Hence we shall focus on gravitational dynamics 
in asymptotically AdS$_5$ spacetime.

Gravitational states dual to shock waves in SYM moving in the $\pm z$ directions 
can be constructed by looking for steady state solutions to Einstein's equations which 
only depend on time through the combination $z_\mp \equiv z \mp t$ \cite{Gubser:2008pc,Grumiller:2008va}.
Consider the Fefferman-Graham coordinate system ansatz for the metric,
\begin{align}
    ds^2 &=  r^2 \big[
	{-} dt^2 + d\bm x^2 + {\textstyle \frac{dr^2}{r^4}}
	+ h_\pm(\bm x_\perp, z_\mp,r) \, dz_\mp^2
    \big] \,,
\label{eq:FG}
\end{align}
where $\bm x \equiv \{x,y,z\}$, $\bm x_\perp \equiv \{x,y\}$ and $r$ is the AdS radial coordinate
with $r = \infty$ the AdS boundary.  \clr{Note that here and in what follows
we set the AdS radius  to unity. }
The anzatz (\ref{eq:FG}) satisfies Einstein's equations provided 
$h_\pm$ satisfy the linear partial differential equation
\begin{equation}
\label{eq:singleshockEin}
\left(\del_\perp^2 +r^4 \partial_r^2 + 5 r^3 \partial_r \right ) h_\pm = 0.
\end{equation}
The solution to (\ref{eq:singleshockEin}) satisfying vanishing  boundary conditions at the AdS boundary is
\begin{align}
    h_\pm(\bm x_\perp,z_\mp,r) &\equiv
     \int \frac{d^2 k}{(2\pi)^2} \>
    e^{i {\bm k} \cdot \bm x_\perp} \,
    \widetilde H_\pm({\bm k},z_\mp) \,
    \frac{8I_2(k/r)}{k^2 r^2} \,,
\label{eq:h}
\end{align}
where $\widetilde H_\pm$ is an arbitrary function.

The metric (\ref{eq:FG}) with $h_\pm$ given by (\ref{eq:h}) consists of a gravitational wave
moving in the $\pm z$ direction --- parallel to the AdS boundary --- at the speed of light.
The fact that $\widetilde H_\pm$ is arbitrary reflects the fact that Einstein's equations 
do not fix the waveform of gravitational waves.  Likewise, in the dual field theory nothing
fixes the waveform of the shock waves.  Indeed, SYM is conformal and contains no bound states
or scales which would otherwise determine the structure of shock waves.  We shall exploit the 
arbitrariness of $\widetilde H_\pm$ to construct states in SYM which superficially at least resemble 
localized and Lorentz contracted protons.  To do so we note that the Fourier transform of 
$\widetilde H_\pm$, $H_\pm$, determines the expectation value of the SYM stress tensor 
via \cite{deHaro:2000vlm}
\begin{equation}
\label{eq:singleshock2}
    \langle T^{00} \rangle = \langle T^{zz}\rangle = \pm \langle T^{0z} \rangle = \frac{N_{\rm c }^2}{2 \pi^2}
    H_\pm(\bm x_\perp,z_\mp),
\end{equation}
with all other components vanishing.  Therefore, once we specify the SYM energy density $\langle T^{00} \rangle$
for a single shock, we specify the dual gravitational wave.  

\begin{figure}[ht!]
\vskip +0.15in
\begin{center}
\includegraphics[scale = 0.58]{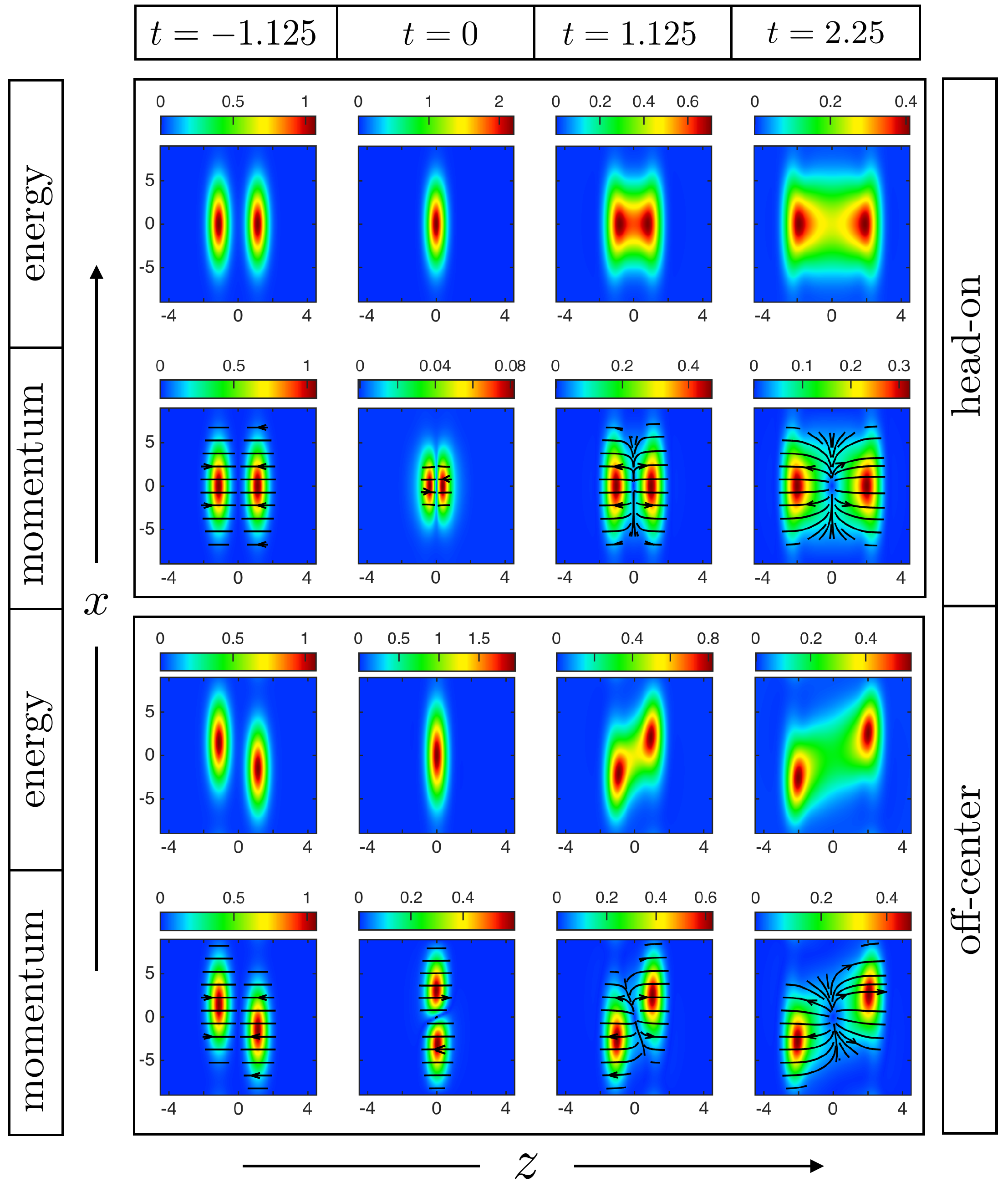}
\end{center}
\vskip -0.1in
\caption{%
The rescaled energy density $\langle \T^{00}\rangle $ 
and rescaled momentum density $|\langle T^{0i} \rangle|$,
for head-on  and off-center collisions at four different times.
Streamlines in the plots of the momentum density
denote the direction of the momentum density.  
At the initial time $t = -1.125$ the ``protons" are at $z = \pm 1.125$.
The non-zero impact parameter $\bm b = 3 \hat x$ is apparent in the off-center collision.
The shocks move in the $\pm z$ direction at the speed of light and collide at $t = z = 0$.
After the collision the remnants of the initial shocks, which remain close to the light cone
$z = \pm t$, are significantly attenuated in amplitude with the extracted
energy deposited inside the light cone.
Note the appearance of transverse flow 
at positive times for both impact parameters.
\label{fig:snapshots}
} 
\end{figure}

We shall employ the single shock energy density 
given in Eq.~(\ref{eq:singleshock}) in the Introduction.  This means
\begin{equation}
H_{\pm}(\bm x_\perp,z_\mp) = \mu^3 
\exp \left [-{\textstyle \frac{1}{2}}( \bm x_\perp \mp {\textstyle \frac{\bm b}{2}})^2/\sigma^2 \right] \delta_w(z_\mp).
\end{equation}
For the smeared $\delta$ function we choose a Gaussian
\begin{equation}
\delta_w(z)=\frac{1}{\sqrt{2 \pi w^2}} e^{-\frac 12 z^2/w^2 },
\end{equation}
with width $\mu w = 0.375$.
Note that evolution inside the future light cone of planar shock collisions with width $\mu w = 0.375$ 
well approximates that of the $\delta$ function limit \cite{Chesler:2015fpa}.  
The shock parameters we employ are given in Eq.~(\ref{eq:shockparms}) in the Introduction.

At early times, $t \ll -w$, the profiles 
$H_\pm$ have negligible overlap and
the precollision geometry can be constructed from (\ref{eq:FG})
by replacing the last term with the sum of corresponding terms from
left and right moving shocks.
The resulting metric satisfies Einstein's equations, at early times,
up to exponentially small errors.

To evolve the precollision geometry forward in time we 
use the characteristic formulation of gravitational dynamics in
asymptotically AdS spacetimes.  Here we simply outline the salient details 
of the characteristic formulation in
asymptotically AdS spacetime.  For details we refer the reader to
\cite{Chesler:2013lia}.
Our metric ansatz reads
\begin{equation}
    ds^2 = r^2 \, g_{\mu \nu}(x,r) \, dx^\mu dx^\nu + 2 \, dr \, dt \,,
\label{eq:ansatz}
\end{equation}
with Greek indices denoting spacetime boundary coordinates, $x^\mu = (t,x,y,z)$.
Near the boundary,
$g_{\mu \nu} = \eta_{\mu \nu}  + g_{\mu \nu}^{(4)}/r^4 + O(1/r^5)$.
The subleading coefficients $g^{(4)}_{\mu\nu}$ determine 
the SYM stress tensor \cite{deHaro:2000vlm}
\begin{equation}
\langle T^{\mu \nu}\rangle = \frac{N_{\rm c}^2}{2 \pi^2} \left [g_{\mu \nu}^{(4)}  + \tfrac{1}{4} \, \eta_{\mu \nu}\, g_{00}^{(4)} \right].
\end{equation}


An important practical matter in numerical relativity is determining the computational domain and excising singularities behind the event horizon.
To excise singularities, we look for the position of an apparent horizon, which 
if it exists must lie behind an event horizon, and stop numerical evolution at its location.
Note that the metric ansatz (\ref{eq:ansatz}) is invariant under the 
residual diffeomorphism 
\begin{equation}
\label{eq:gaugetrans}
r \to r+ \lambda(t,\bm x),
\end{equation}
for arbitrary $\lambda$.  If the apparent horizon 
has planar topology and lies at say $r = r_{h}(t,\bm x)$, then we may use this residual diffeomorphism invariance to set 
the apparent horizon to be at $r = 1$ by choosing $\lambda = -1 + r_h$.  
Let $\nabla$ be the covariant derivative under the spatial metric $g_{ij}$.
Fixing the apparent horizon to be at constant $r$, at the horizon the metric must satisfy \cite{Chesler:2013lia}
%
\begin{equation}
\label{eq:horizonfix}
\left(\partial_t - {\textstyle \frac{1}{2}}g_{00} \partial_r \right) [\det g_{ij} ]^{1/6}  + {\textstyle \frac{1}{2}}  (\partial_r [\det g_{ij} ]^{1/6} ) g^{ij} g_{0i} g_{0j} - {\textstyle \frac{1}{3}} [\det g_{ij} ]^{1/6} g^{ij} \nabla_i g_{0 j} = 0.
\end{equation}
We note that Eq.~(\ref{eq:horizonfix}) is a constraint on initial data: if (\ref{eq:horizonfix}) is satisfied at any fixed time
and appropriate boundary conditions are enforced, then Einstein's equations guarantee (\ref{eq:horizonfix}) remains satisfied at future times.  
See \cite{Chesler:2013lia} for more details. Hence, we only need to enforce (\ref{eq:horizonfix}) on initial data.
For given initial data we compute the LHS of (\ref{eq:horizonfix}) at $r = 1$ and search for a radial shift $\lambda$ such that 
(\ref{eq:horizonfix}) is satisfied to some chosen accuracy.  The shift $\lambda$ is computed using Newton's method.  
In our numerical simulations we demand that the norm of the LHS of (\ref{eq:horizonfix}) is $5 \times 10^{-3}$ or smaller.

To generate initial data for our characteristic evolution,
we numerically transform the precollision metric in
Fefferman-Graham coordinates 
to the metric ansatz (\ref{eq:ansatz}).  
We begin time evolution at $t = -1.5$.
For both head-on and off-center initial data we find an apparent horizon at time $t = -1.5$ ---
\textit{before} the collision on the boundary.  Since the apparent horizon must lie inside an event horizon, 
this means that there is an event horizon even before the collision takes place on the boundary.
Why must an event horizon exist before the collision?
Empty AdS contains a cosmological horizon, which is simply the Poincare horizon.  It takes a 
light ray an infinite amount of (boundary) time to travel from the Poincare horizon to any bulk point in the geometry.
This means that if a planar topology event horizon exists at any time --- say after the collision --- then it must have also existed at all times in the past 
and coincided with the Poincare horizon at $t = -\infty$.  Similar conclusions were reached within the context 
of a quench studied in \cite{Chesler:2008hg}.  
 
For numerical evolution we periodically compactly spatial directions with transverse size $L_x = L_y  = 18$
and longitudinal size $L_z = 9$ and and evolve from $t = -1.5$ to $t = 2.5$.
For the head-on collision, evolution is performed using a 
spectral grid of size $N_x = N_y = 39$, $N_z = 155$ and $N_r = 48$
while for the off-center collision we increase the number of transverse points to 
$N_x = N_y = 45.$%
\footnote
  {
   We also modify the initial data by adding a small uniform background energy density,
   equal to 0.5\% and 1\% of the peak energy density of the incoming shocks
   for the head-on and off-center collisions, respectively.
   This allows use of a coarser grid, reducing memory requirements and increasing computational speed.
  } 
To monitor the convergence of our numerical solution, we monitor violations 
of the horizon fixing condition (\ref{eq:horizonfix}).   
Since in the continuum limit (\ref{eq:horizonfix}) is an initial value 
constraint, violations of (\ref{eq:horizonfix}) after the initialization time $t = -1.5$
provide a simple measure of the degree in which our numerics are converging to the continuum limit.
For both head-on and off-center collisions violations of (\ref{eq:horizonfix}) are $5 \times 10^{-3}$ or smaller
at all times, which is the same size of the allowed violations of (\ref{eq:horizonfix}) at the initialization time. 
The fact that (\ref{eq:horizonfix}) remains satisfied after the initialization time provides 
a nontrivial check of the convergence of our numerics.

\section{Results}

\begin{figure}[ht]
\vskip +0.15in
\begin{center}
\includegraphics[scale = 0.43]{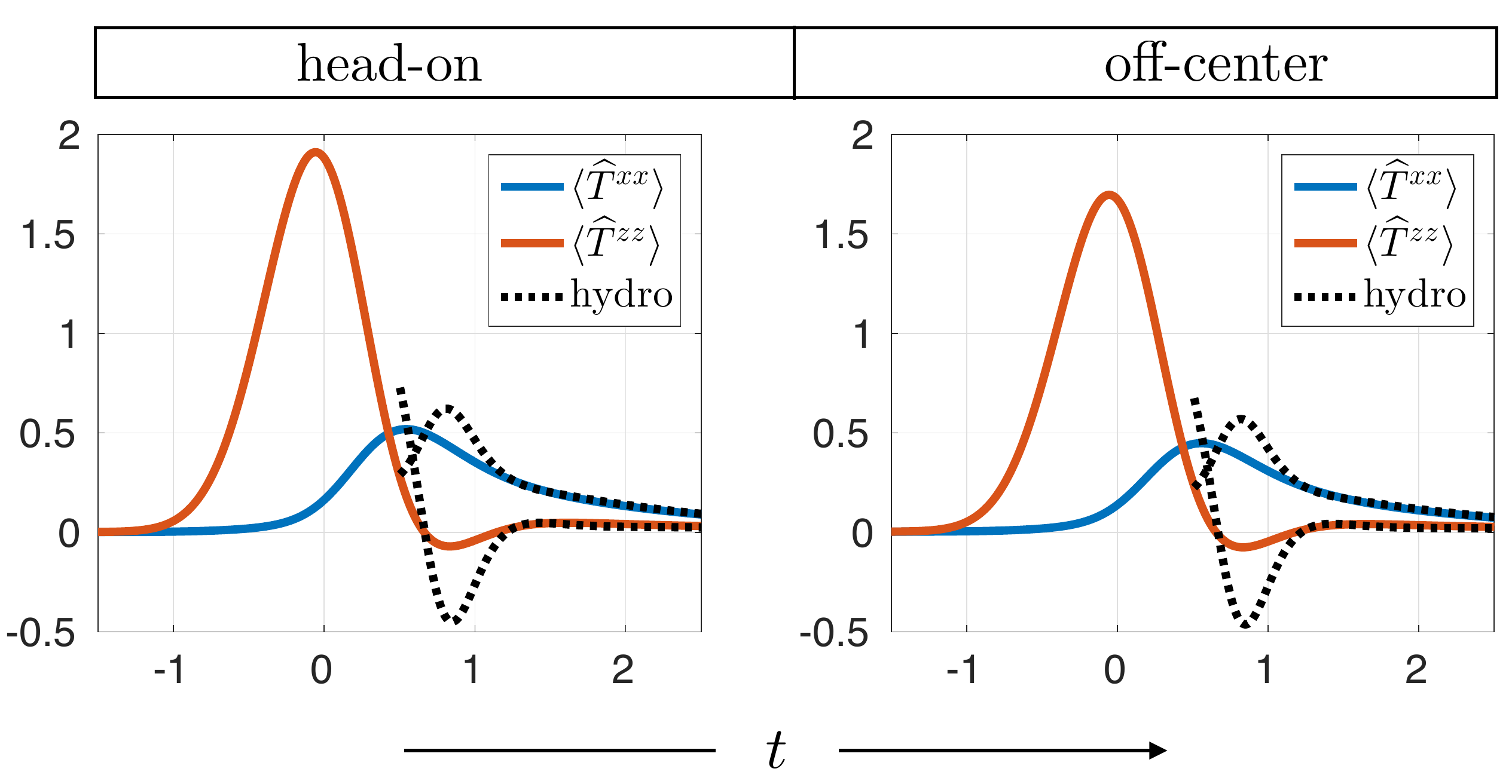}
\end{center}
\vskip -0.1in
\caption{%
Rescaled stress tensor components $\langle \T^{xx} \rangle$ and $\langle \T^{zz} \rangle$ 
at the spatial origin, $x = y = z = 0$, as a function of time for head-on and off-center collisions.  Dashed lines denote 
the hydrodynamic approximation (\ref{eq:hydrocons}).  Note that the stress in the different collisions are nearly identical.
Around $t = 0$ the systems are highly anisotropic and far from equilibrium.  
Nevertheless, at this point in space, the systems begins to 
evolve hydrodynamically at $t = t_{\rm hydro} \approx1.2$.  Note that the large pressure anisotropy
persists during the hydrodynamic evolution.
\label{fig:pressures}
} 
\end{figure}

\label{sec:results}

Define the rescaled stress
\begin{equation}
\langle \T^{\mu \nu} \rangle \equiv \frac{2 \pi^2}{N_{\rm c}^2} \langle T^{\mu \nu} \rangle.
\end{equation}
In Fig.~\ref{fig:snapshots} we plot the rescaled energy density $\langle \T^{00}\rangle$ 
and the rescaled momentum density $\langle \T^{0i} \rangle$
in the plane $y=0$ for head-on and off-center collisions at several values of 
time.  The color scaling in the plots of $\langle \T^{0i}\rangle$ denotes the magnitude of the momentum
density $|\langle \T^{0i}\rangle|$, while the streamlines indicate the direction 
of $\langle \T^{0i}\rangle$.  At time $t = -1.125$ the systems consist of two separated
``protons" centered on $z = \pm 1.125$, $x = 0$ for the head-on collision and $z = \pm 1.125$, $x = \mp \frac{b}{2}$ 
for the off-center collision.  In both collisions the ``protons" move towards each other at the speed of light
and collide at $t = z = 0$.  Note that at $t = 0$ the momentum density 
of the head-on collision nearly cancels out.  Indeed, at $t = 0$ the
energy and momentum densities for both collisions are at the order 10\% level just the linear superposition 
of that of the incoming ``protons."  This indicates that nonlinear interactions 
haven't yet modified the stress.  Similar observations were made within the context of 
planar shock collisions in \cite{Casalderrey-Solana:2013aba}.  Nevertheless, at subsequent times
we see that the collisions dramatically alter the outgoing state.  For both collisions the amplitude of the energy and momentum
densities near the forward light cone have decreased by order 50\% relative to the past light cone, with the 
lost energy being deposited inside the forward light cone.  Moreover, after $t=0$ we see the presence of transverse
flow for both impact parameters.  As we demonstrate below, the evolution of the postcollision debris for both impact parameters is well described by viscous hydrodynamics.

\begin{figure}[h]
\vskip +0.15in
\begin{center}
\includegraphics[scale = 0.5]{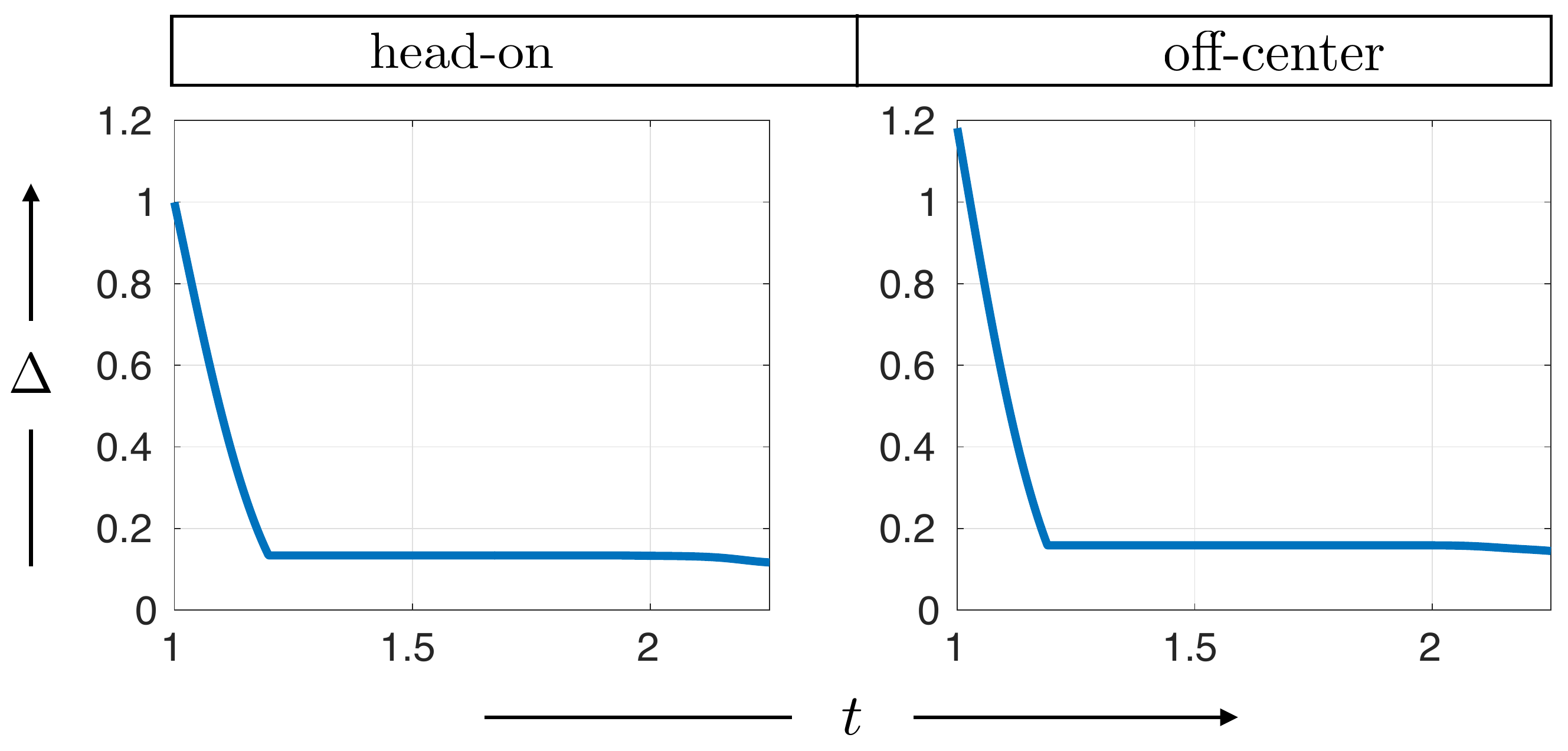}
\end{center}
\vskip -0.1in
\caption{%
The hydrodynamic residual $\Delta$ at 
at the spatial origin, $x = y = z = 0$, as a function of time for head-on and off-center collisions.  
Note that $\Delta$ need not be differentiable due to the max and matrix norm in its definition (\ref{eq:Delta}).  
Regions with $\Delta \ll 1$ have hydrodynamized.  Prior to $t = t_{\rm hydro} \approx 1.2$,
the residual is large and hydrodynamics is not a good description of the evolution.  However, as $t \to t_{\rm hydro}$ 
the residual dramatically decreases.  Thereafter, $\Delta$ continues to decrease but at a much slower rate.
The rapid decay of $\Delta$ prior to $t = t_{\rm hydro}$ reflects the rapid decay of nonhydrodynamic modes in the 
far-from-equilibrium state created by the collisions.
\label{fig:hydroresorigin}
} 
\end{figure}

In Fig.~\ref{fig:pressures} we plot $\langle  \T^{xx} \rangle$ and $\langle \T^{zz} \rangle$ and the 
hydrodynamic approximation $ \T^{xx}_{\rm hydro}$ and $ \T^{zz}_{\rm hydro}$
at the origin $\bm x = 0$ for both impact parameters.  
At this point the stress tensor is diagonal, the flow velocity $\bm u = 0$, and 
$\langle \T^{xx} \rangle$ and $\langle \T^{zz}\rangle$ 
are simply the pressures in the $x$ and $z$ directions.  
As shown in the figure, the pressures start off at zero before the collision.  
Note that the pressures are remarkably similar for both collisions.  Evidently, the finite
impact parameter employed here has little effect on the pressures at the origin.
During the collisions  the pressures increase dramatically, 
reflecting a system which is highly anisotropic and far from equilibrium.  Nevertheless, after time 
\begin{equation}
t = t_{\rm hydro} \approx 1.2,
\end{equation}
the system has hydrodynamized at the origin and 
the evolution of the pressures for both impact parameters 
is well described by the hydrodynamic constitutive relations.  Remarkably, at this time 
$ \langle \T^{xx} \rangle$ is order 10 times greater than  $ \langle \T^{zz} \rangle$.  

\begin{figure}[h]
\vskip +0.15in
\begin{center}
\includegraphics[scale = 0.45]{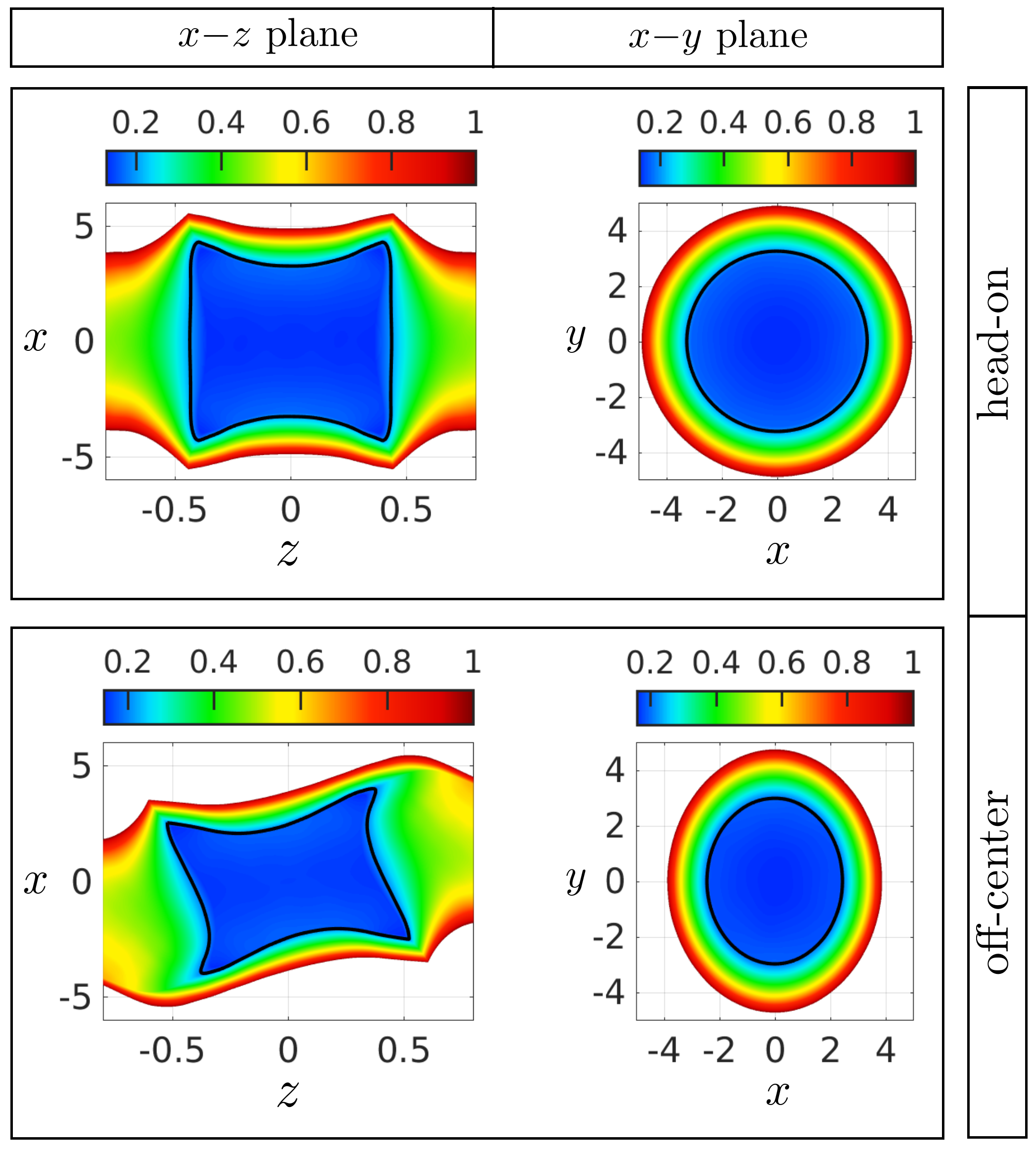}
\end{center}
\vskip -0.1in
\caption{%
The hydrodynamic residual $\Delta$ in the $x{-}z$ and $x{-}y$ planes at time $t = 1.25 \,t_{\rm hydro} = 1.5$
for both head-on and off-center collisions.  Regions with $\Delta \ll 1$ have hydrodynamized.
Note that we have restricted the plot to $\Delta < 1$ in order to highlight the hydrodynamic behavior.
The black curve in the plots is the surface $\Delta = 0.2$.
For both impact parameters there is a crisply defined region --- whose boundary is well approximated by 
the $\Delta = 0.2$ surface --- 
where $\Delta \ll 1$.
We identify the matter in the interior of the $\Delta = 0.2$ surface
as a droplet of liquid.  Outside the $\Delta = 0.2$ surface
$\Delta$ rapidly increases, indicating the presence of nonhydrodynamic modes on the surface 
of the droplet.  Note the irregularity in the off-center collision droplet shape in the $x{-}z$ plane  
is due to nonhydrodynamic modes and not fluid rotation
in the $x{-}z$ plane.
For the head-on collision the $\Delta = 0.2$ surface is circular in the 
$x{-}y$ plane.  In contrast, for the off-center collision the $\Delta = 0.2$ surface is elliptical in the 
$x{-}y$ plane, with the the short axis of the ellipse oriented in the same direction as the impact parameter $\bm b = 3 \hat x$.
Nevertheless, for both collisions the transverse radius of the $\Delta = 0.2$ surface is roughly the same 
and equal to $R \sim 3$, which is just the radius $\sigma$ of our ``protons."    
\label{fig:hydrores}
} 
\end{figure}

In Fig.~\ref{fig:hydroresorigin} we plot the hydrodynamic residual $\Delta$
at $\bm x = 0$ for both impact parameters.  We see from the figure that $\Delta$ is large at times $t < t_{\rm hydro}$.  
However, as $t \to t_{\rm hydro}$ the residual $\Delta$ rapidly decays to $\Delta \lesssim 0.2$.
After $t = t_{\rm hydro}$ the residual continues to decay, but at a much slower rate than 
before $t = t_{\rm hydro}$.
The rapid decay of $\Delta$ prior to $t = t_{\rm hydro}$ indicates the rapid decay of nonhydrodynamic modes.  

Over what region of space has the system hydrodynamized?
To study the spatial domain of applicability of hydrodynamics, 
in Fig.~\ref{fig:hydrores} we plot $\Delta$ in the $x{-}z$ and $x{-}y$ planes at time $t = 1.25 \, t_{\rm hydro} = 1.5$
for both head-on and off-center collisions.  
In order to highlight the hydrodynamic behavior, we omit regions where $\Delta > 1$.
For both collisions we see a region where $\Delta \ll 1$.  
In this region the stress has hydrodynamized and correspondingly, we identify the matter in the interior
as a droplet of liquid.  Also included in the figure is the surface $\Delta = 0.2$, shown
as the solid curve.  For both collisions $\Delta$ increases dramatically outside the 
$\Delta = 0.2$ surface, indicating the presence of nonhydrodynamic modes.
In what follows we shall use the $\Delta = 0.2$ surface to define the surface of our droplet of
liquid.   

Note the irregularity in the off-center collision droplet shape in the $x{-}z$ plane  
is due to nonhydrodynamic modes and not fluid rotation
in the $x{-}z$ plane.  Indeed, in the interior of the droplet the vorticity is small with 
$|| \Omega^{\mu \nu} || \sim 0.1 || \sigma^{\mu \nu} ||$.
Additionally, note the difference in droplet shape in the $x{-}y$ plane for head-on and off-center collisions.
For the head-on collision droplet's surface is circular in the 
$x{-}y$ plane.  In contrast, for the off-center collision the droplet's surface is elliptical in the 
$x{-}y$ plane, with the the short axis of the ellipse oriented in the same direction as the impact parameter $\bm b = 3 \hat x$.
Nevertheless, for both collisions the transverse radius of the droplet is roughly the same 
and equal to $R \sim 3$, which is just the radius $\sigma$ of our ``protons" employed in (\ref{eq:shockparms}).

\begin{figure}[ht]
\vskip +0.15in
\begin{center}
\includegraphics[scale = 0.45]{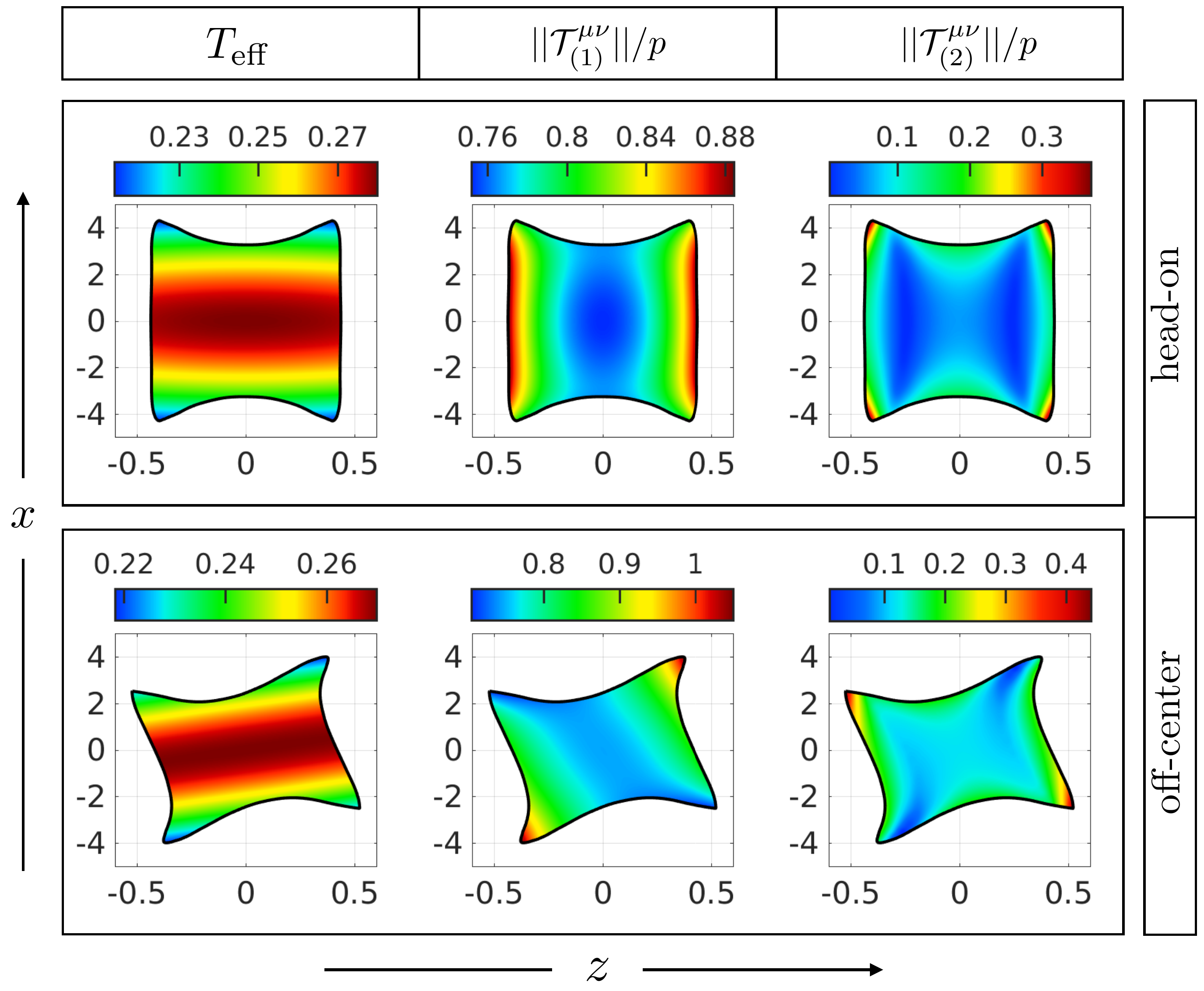}
\end{center}
\vskip -0.1in
\caption{%
The effective temperature $T_{\rm eff}$ and the first and second order gradient norms $||\mathcal T_{(1)}^{\mu \nu}||$
and $||\mathcal T_{(2)}^{\mu \nu}||$ in units of the average pressure $p$ for head-on and off-center collisions.
All plots are shown at time $t = 1.25 \, t_{\rm hydro} = 1.5$ and are restricted to the domain inside the $\Delta = 0.2$ surface shown in Fig.~\ref{fig:hydrores}, which is also shown here as the solid curve.  In this region hydrodynamics is a good approximation to the evolution of the stress.
For both collisions the average effective temperature in the displayed region is $T_{\rm eff} \sim 0.25$.
Inside the $\Delta = 0.2$ surface $||\mathcal T_{(2)}^{\mu \nu}|| \ll ||\mathcal T_{(1)}^{\mu \nu}||$, meaning 
second order gradient corrections are negligible.  
Note however, that 
near the surface of the droplet, where nonhydrodynamic modes are excited, $||\mathcal T_{(2)}^{\mu \nu}||$ begins to become comparable to 
$||\mathcal T_{(1)}^{\mu \nu}||$.
\label{fig:fluidplots}
} 
\end{figure}

\begin{figure}[ht]
\vskip +0.15in
\begin{center}
\includegraphics[scale = 0.45]{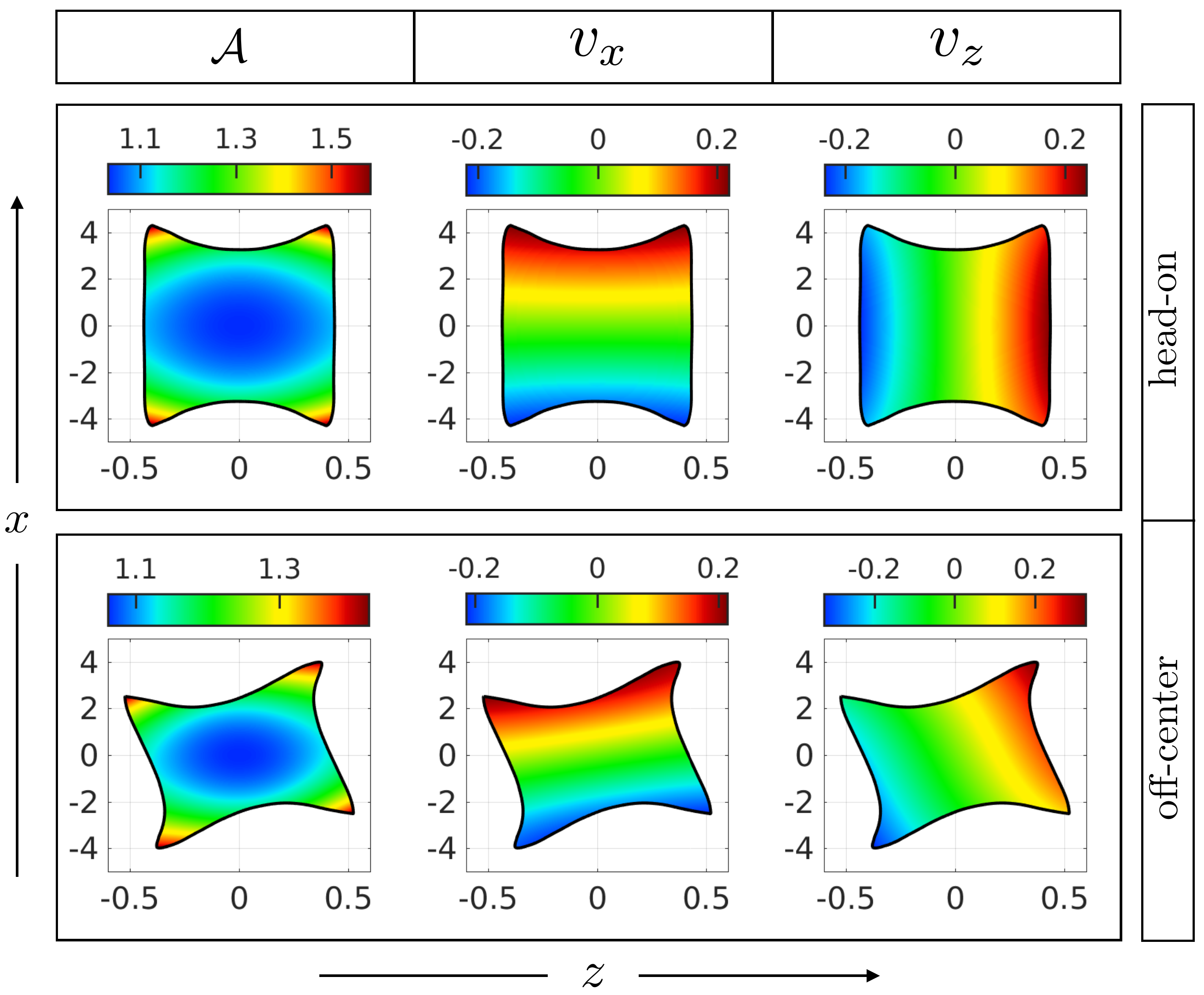}
\end{center}
\vskip -0.1in
\caption{%
The anisotropy $\mathcal A$ and the transverse and longitudinal components of the 
fluid 3-velocity $\bm v \equiv \bm u/u^0$ for head-on and off-center collisions.
All plots are shown at time $t = 1.25 \, t_{\rm hydro} = 1.5$ and are restricted to the domain inside the $\Delta = 0.2$ contour shown in Fig.~\ref{fig:hydrores}, which is also shown here as the solid curve.  In this region hydrodynamics is a good approximation of the evolution of the stress.
In the limit of ideal hydrodynamics the anisotropy vanishes.  The order 1 anisotropy observed here for both collisions 
reflects the fact that there are large gradients in the system.  Likewise, for both collisions we see that $v_x$ and $v_z$ are similar in magnitude.  The large transverse velocity is also a signature of large gradients.
\label{fig:anisoandvelplots}
} 
\end{figure}

In Figs.~\ref{fig:fluidplots}-\ref{fig:fluidvelxy} 
we restrict our attention to time $t = 1.25 \,t_{\rm hydro} = 1.5$ and to the region inside the 
droplet of fluid.  We explain the 
coloring of the different plots below.  In the left column of Fig.~\ref{fig:fluidplots} we show the 
effective temperature $T_{\rm eff}$, defined 
in Eq.~(\ref{eq:transcoeffs}).
From the figure it is evident that the average value of the temperature inside the droplet
is $T_{\rm eff} \sim 0.25$ for both collisions.
We therefore obtain the dimensionless measure
of the transverse size of the droplet,
\begin{equation}
R T_{\rm eff} \sim 1.
\end{equation}
Similar conclusions were reached in \cite{Chesler:2015bba}, 
where a holographic model of proton-nucleus collisions was studied.
Additionally, note 
\begin{equation}
t_{\rm hydro} T_{\rm eff} \sim 0.3,
\end{equation}
indicating rapid hydrodynamization.
Similar hydrodynamization times were observed in \cite{Chesler:2015bba,Casalderrey-Solana:2013aba,Heller:2013fn,Chesler:2009cy}. We therefore conclude that hydrodynamic evolution
applies even when both the system size and time after the collision are 
on the order of microscopic scale $1/T_{\rm eff}$. 

In the middle and right columns of Figs.~\ref{fig:fluidplots} we plot the size of first and second order gradient 
corrections to the hydrodynamic constitutive relations, $|| \mathcal T^{\mu \nu}_{(1)}||$
and $|| \mathcal T^{\mu \nu}_{(2)}||$, normalized by the average pressure $p$.  
In the interior of the droplet $|| \mathcal T^{\mu \nu}_{(2)}||$ is nearly an order of magnitude smaller than  
$|| \mathcal T^{\mu \nu}_{(1)}||$ for both impact parameters, meaning that second order gradient 
corrections are negligible.%
\footnote
  {
  We note however, that inside the $\Delta = 0.2$ surface shown in Fig.~\ref{fig:hydrores} we have
  $|| \langle T^{\mu \nu} \rangle - T^{\mu \nu}_{\rm hydro} || \sim ||\mathcal T^{\mu \nu}_{(2)}||$.
  Given that  $T^{\mu \nu}_{\rm hydro}$ is computed to second order in gradients, this could mean that 
  third order gradient corrections to $T^{\mu \nu}_{\rm hydro}$ are comparable to the second order gradient corrections.   
  Alternatively, and in our opinion more likely, this could reflect the fact that the spectrum of the discretized Einstein 
  equations we solved is different from that of the continuum limit used to obtain the gradient expansion of 
  $T^{\mu \nu}_{\rm hydro}$.  This difference must manifest itself at suitably high order in gradients and can be ameliorated by 
  using a finer discretization scheme.
  }

\section{Discussion}
\label{sec:discuss}

Given that $1/T_{\rm eff}$ is the salient microscopic scale in strongly coupled 
plasma --- akin to the mean free path at weak coupling --- it is remarkable that hydrodynamics
can describe the evolution of systems as small at $1/T_{\rm eff}$.  By setting the ``proton" radius 
equal to the actual proton radius, $\sigma \sim 1$ fm, and using $N_{\rm c} = 3$ colors, the 
single shock energy (\ref{eq:shockenergy}) is $\sim 20$ GeV.  Likewise, the effective temperature of the produced plasma is
$T_{\rm eff} \sim 200$ \clr{MeV}.  Given that collisions at RHIC and the LHC have higher energies and 
temperatures, it is natural to expect $R T_{\rm eff}$ to be larger in RHIC and LHC collisions 
than the simulated collision presented here.  
We therefore conclude that droplets of liquid the size of the proton need not be thought of as unnaturally small.  
Evidently, there are theories --- namely strongly coupled SYM --- which enjoy hydrodynamic evolution in even smaller systems.
%
%
%
%
%

While hydrodynamics is a good description of the evolution of our collision, it should be noted that the
produced hydrodynamic flow has an extreme character to it.  In particular, 
when the system size is on the order of the microscopic scale $1/T_{\rm eff}$, 
gradients must be large.  To quantify the size of gradients we introduce 
the anisotropy function 
\begin{equation}
\label{eq:aniso}
\mathcal A \equiv \frac{{\rm max}(p_{(i)}) - {\rm min}(p_{(i)})}{p},
\end{equation}
where the pressures $p_{(i)}$ are determined by the eigenvalue equation (\ref{eq:veldef})
and again $p$ is the average pressure, $p = \frac{\epsilon}{3} = {\rm avg}(p_{(i)})$.  In ideal hydrodynamics, where 
all the pressures $p_{(i)}$ are equal, the anisotropy vanishes.  It therefore follows that 
after the system has hydrodynamized, any nonzero anisotropy must be due to gradient corrections 
in the hydrodynamic constitutive relations (\ref{eq:hydrocons}).  In the left
column of Fig.~\ref{fig:anisoandvelplots}
we plot $\mathcal A$ at time $t = 1.25 \, t_{\rm hydro} = 1.5$ in the interior of the droplet.  
For both impact parameters $\mathcal A \sim 1$, with larger anisotropies near the droplet's surface.  
Evidently, gradients are large and ideal hydrodynamics is not a good approximation.  This is also evident in the pressures displayed in Fig.~\ref{fig:pressures}, where the transverse and longitudinal pressures differ by a factor of 5 to 10. 
Moreover, as shown in Fig.~\ref{fig:fluidplots}, 
$||\mathcal T^{\mu \nu}_{(1)}|| \sim p$, so the first order gradient correction $\mathcal T^{\mu \nu}_{(1)}$ is 
comparable to the ideal hydrodynamic stress.   However, despite the presence of large gradients, 
the above observed large anisotropy must be almost entirely due to the first order gradient correction 
$\mathcal T^{\mu \nu}_{(1)}$ alone, since $|| \mathcal T^{\mu \nu}_{(2)}||$ is nearly 
an order of magnitude smaller than $|| \mathcal T^{\mu \nu}_{(1)}||$ inside the droplet of fluid.
This means that viscous hydrodynamics alone is sufficient to capture the 
observed large anisotropies and evolution of the system.

It is remarkable that the second order gradient correction $\mathcal T^{\mu \nu}_{(2)}$
is negligible even when gradients are large enough to produce an order 1 anisotropy.  
It is also remarkable that the presence of large gradients does not substantially excite nonhydrodynamic modes 
(\textit{e.g.} nonhydrodynamic quasinormal modes).  In an infinite box of liquid, 
nonhydrodynamic modes decay over a timescale of order
$1/T_{\rm eff}$.  Evidently, in the interior of the small droplet studied here, 
the rate of decay of nonhydrodynamic modes is stronger than the rate in which they are excited.

\begin{figure}[ht]
\vskip +0.15in
\begin{center}
\includegraphics[scale = 0.45]{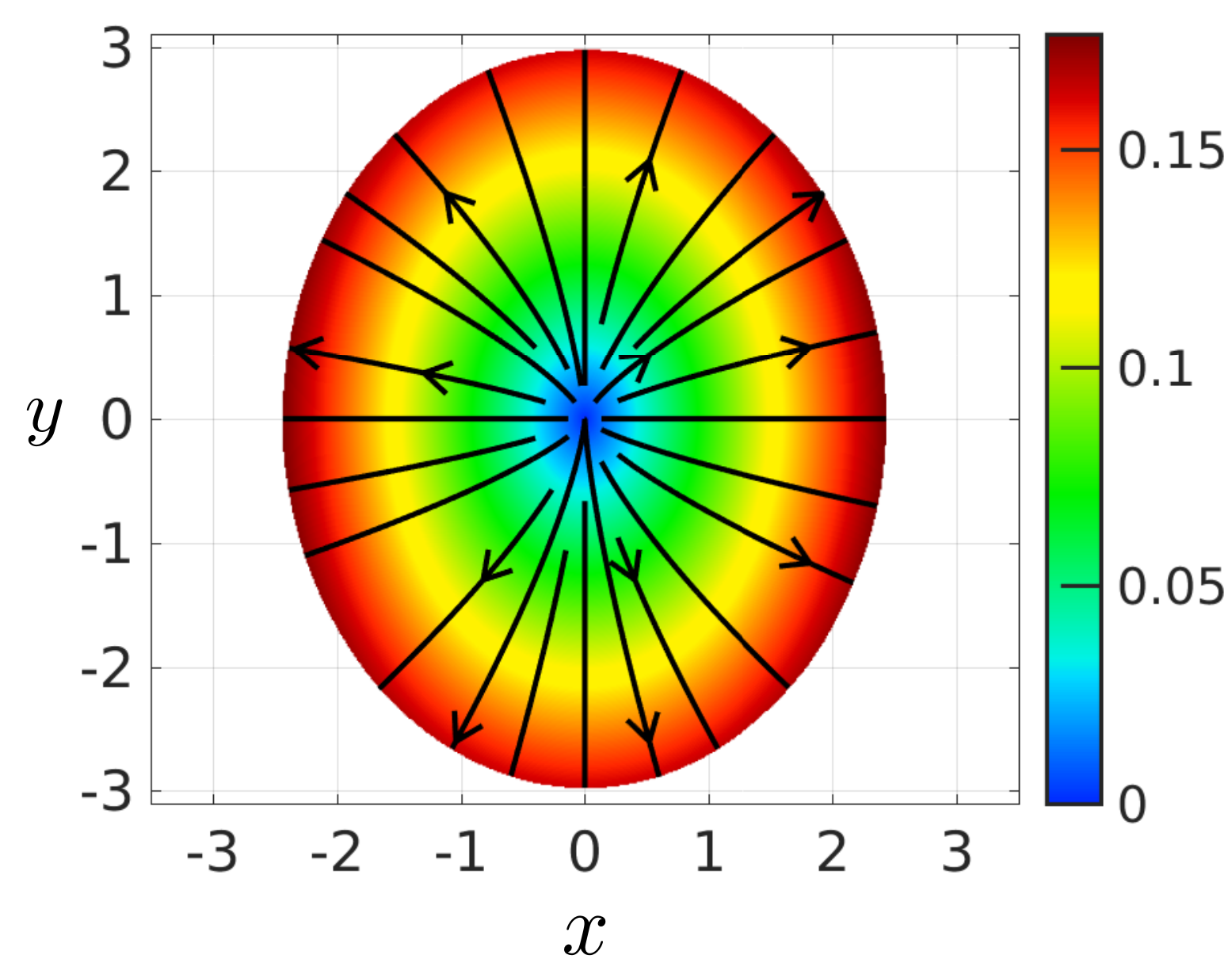}
\end{center}
\vskip -0.1in
\caption{%
The fluid 3-velocity $\bm v$ in the $x{-}y$ plane at time $t = 1.25 t_{\rm hydro}$.  The coloring denotes the 
magnitude of the 3-velocity while the streamlines denote its direction.  Note that we have restricted the plot to the 
interior of the $\Delta = 0.2$ surface.  
\label{fig:fluidvelxy}
} 
\end{figure}

We now turn to the nature of the produced hydrodynamic flow.
Also shown in Fig.~\ref{fig:anisoandvelplots} are the transverse and longitudinal 
components of the fluid 3-velocity, $\bm v \equiv \bm u/u^0$, at time 
$t = 1.25 \, t_{\rm hydro} $ for both head-on and off-center collisions.
$v_x$ and $v_z$ are similar in magnitude for both impact parameters.  
Given that the transverse velocity starts from zero and is driven by transverse gradients, 
the rapid development of transverse flow observed here must
reflect the presence of large transverse gradients.  In Fig.~\ref{fig:fluidvelxy} we plot 
the 3-velocity in the $x{-}y$ plane at time 
$t = 1.25 \, t_{\rm hydro} $ in the interior of the droplet produced in the off-center collision.  The coloring in the plot
denotes the magnitude of the fluid 3-velocity while the streamlines denote its direction.
The fluid velocity is not radial, which is natural given the finite impact parameter
and the elliptical shape of the droplet in the $x{-}y$ plane.
However, the flow of \textit{momentum} is nearly radial.  To quantify this, define the 
Fourier coefficient $c_n \equiv \int d^3 x \langle T^{0 i} \rangle \hat \rho^i \cos n \phi$ 
where $\hat \rho = \cos \phi \hat x + \sin \phi \hat y$ is the radial unit vector with $\phi$ the azimuthal
angle.  Deviations in radial flow are encoded in $c_2$.  However, at time 
$t = 1.25 \, t_{\rm hydro}$, the coefficient $c_2$ is only 1\% of $c_0$, indicating small deviations
from radial momentum flow.
It would be interesting to see how $c_2$ grows as time progresses.  It would also be interesting to 
see how $c_2$ is affected by different choices of shock profiles.
Indeed, off-center collisions of 
Gaussian shock profiles with large transverse width $\sigma$ 
generate purely radial flow \cite{Chesler:2015wra}, since the overlap 
of two off-center Gaussians is still a Gaussian and thus rotationally invariant about the $z$ axis.

Let us now ask how the droplet of fluid produced in the collision further evolves at later times.
In conformal SYM the droplet must expand forever with size $R \sim t$.  This is very different from a confining 
theory, where the droplet cannot expand indefinitely.  Likewise, as the droplet expands it must cool.  
Its easy to reason from energy conservation that the temperature must 
decrease slower than $1/t$, meaning that $R T_{\rm eff} \to \infty$ as $t \to \infty$, so 
hydrodynamics becomes a better and better description of the evolution.  We expect the surface
of the droplet to become smoother and smoother as time progresses.  Likewise, we expect 
nonhydrodynamic modes near the surface of the droplet to continue to decay with the fraction 
of energy behaving hydrodynamically approaching 100\% as $t \to \infty$.   Indeed, such 
behavior was observed within the context of planar shock collisions \cite{Chesler:2015fpa}, where 
nonhydrodynamic modes near the light cone decayed, with the lost energy transported inside the light cone
where it hydrodynamizes.

It would be interesting to push our analysis further.  How big are the smallest drops of liquid?
Clearly $R T_{\rm eff}$ cannot be made arbitrarily small since the hydrodynamic gradient expansion (\ref{eq:hydrocons})
breaks down when $R T_{\rm eff} \ll 1$. Moreover, in the gravitational description there likely exists a critical energy $E_c$ 
(or alternatively a critical impact parameter $b_c$ \cite{Lin:2009pn}) below (above) which no black hole is formed and which for $E = E_c + 0^+  $(or $b = b_c + 0^-)$ critical gravitational collapse occurs \cite{Choptuik:1992jv}.  The absence of a black hole when $E < E_c$ (or $b > b_c$) means that in the dual field theory the collisional debris will not evolve hydrodynamically at any future time.  Clearly it would be interesting to look for criticality and the associated signatures of hydrodynamics turning off.  

It would also be of great interest to study collisions in confining theories.  Aside from being more 
realistic, collisions in confining theories have the added feature that the produced droplet --- a plasma ball ---
cannot expand indefinitely.  As the system cools it must eventually reach the confinement/deconfinement transition
and freeze into a gas of hadrons.  In the large $N_{\rm c}$ limit freezeout is suppressed and 
plasma balls are metastable with a lifetime of order $N_{\rm c}^2$ \cite{Aharony:2005bm}.
At $N_{\rm c} =\infty$, plasma balls must eventually equilibrate and become static due to internal frictional
forces.  Therefore, in a large $N_{\rm c}$ confining theory the question of how big are the smallest drops 
of quark-gluon plasma is tantamount to how big are the smallest plasma balls.  An illuminating 
warmup problem is that of SYM on a three sphere of radius $R$.  In the large $N_{\rm c}$ 
limit the theory becomes confining \cite{Witten:1998zw,Aharony:2003sx,Aharony:2005bq}.
Therefore, instead of studying plasma balls produced in a collision, one can study 
plasma confined to the three sphere and analyze the plasma's behavior as the radius of the sphere is changed.

The dual gravitational description of SYM on a three sphere is 
that of gravity in global AdS$_5$$\times S^5$.
At strong coupling, and in the canonical ensemble, the deconfinement
temperature $T_c$ is 
\begin{equation}
\label{eq:criticaltemp}
R T_c = \frac{3}{2 \pi} \approx 0.48,
\end{equation}
with the deconfinement transition manifesting itself in the gravity description as the Hawking-Page phase transition \cite{Witten:1998zw}.  \clr{Above the deconfinement transition the dual geometry contains a black hole, the boundary 
energy density is order $N_{\rm c}^2$, and the boundary state is that of a quark-gluon plasma with hydrodynamic evolution. Below the transition the dual geometry is just thermal AdS and the boundary state has order $N_{\rm c}^0$ energy and does not admit a hydrodynamic description.  Hence in the canonical ensemble of SYM living on a three sphere, the answer to the question of how big are the smallest drops of quark-gluon plasma is given by (\ref{eq:criticaltemp}).}

The situation is less clear in the microcannonical ensemble.  
In this case the dual black hole geometry experiences a Gregory-Laflamme instability \cite{Gregory:1993vy}
on the $S^5$ \cite{Hubeny:2002xn,Buchel:2015gxa,Dias:2015pda}.  The instability sets in
at 
\begin{equation}
\label{eq:gl}
R T_{\rm eff} \approx 0.50.
\end{equation}
It is unknown what the final state of the Gregory-Laflamme instability is.  
One possibility is that 
there do not exist stable black hole solutions below (\ref{eq:gl}).  In this case the instability 
could go on indefinitely, perhaps generating structures on the order of the Planck scale
where quantum effects will be important.  Another possibility is that there exists equilibrium black
holes with broken rotational invariance on the $S^5$.  In this case the final state of the 
Gregory-Laflamme instability would asymptote to this solution.
In either case, it seems likely that the onset of the Gregory-Laflamme instability destroys hydrodynamic
evolution in the dual field theory.  It therefore seems that in either ensemble, the smallest drops of quark-gluon plasma
in the confining phase of SYM have size $R T_{\rm eff} \approx 1/2.$

\section{Acknowledgments}
This work is supported by the Fundamental Laws Initiative at Harvard.  I thank Krishna Rajagopal, Andy Strominger and Larry Yaffe for useful discussions.

\bibliographystyle{JHEP}
\bibliography{refs}
\end		{document}